# Pseudospin-valley-coupled phononic topological insulator with edge and corner sates

Haiyan Fan, Baizhan Xia[*], Shengjie Zheng, Liang Tong

State Key Laboratory of Advanced Design and Manufacturing for Vehicle Body,

Hunan University, Changsha, Hunan, People's Republic of China, 410082

* Correspondence to: xiabz2013@hnu.edu.cn.

**Topologically protected gapless edge states are phases of quantum matter which behave as massless Dirac fermions, immunizing against disorders and continuous perturbations. Recently, a new class of topological insulators (TIs) with topological corner states have been theoretically predicted in electric systems, and experimentally realized in two-dimensional (2D) mechanical and electromagnetic systems, electrical circuits, optical and sonic crystals, and elastic phononic plates. Here, we demonstrate a pseudospin-valley-coupled phononic TI, which simultaneously exhibits gapped edge states and topological corner states. Pseudospin-orbit coupling edge states and valley-polarized edge state are respectively induced by the lattice deformation and the symmetry breaking. When both of them coexist, these topological edge states will be greatly gapped and the topological corner state emerges. Under direct field measurements, the robust edge propagation behaving as an elastic waveguide and the topological corner mode working as a robust localized resonance are experimentally confirmed. The pseudospin-valley coupling in our phononic TIs can be well-controlled which provides a reconfigurable platform for the multiple edge and corner states, and**

**exhibits well applications in the topological elastic energy recovery and the highly sensitive sensing.**

The topological phases of matter, primarily discovered in electronic systems, have significantly renewed our understanding of condensed matter physics, and recently have extended to classical wave systems, from optical and electromagnetic systems[1-20] to acoustic and elastic phononic systems[21-32]. The key property of the two-dimensional (2D) topological insulators (TIs) is the topologically protected one-dimensional (1D) gapless edge states which immunize against disorders and perturbations. Generally, there are two different mechanisms for the generation of topological edge states in time-reversal invariant systems. The first one is the lattice deformation, based on which the double Dirac cone generated at Brillouin zone (BZ) center will open a bandgap. This bandgap is topologically nontrivial as the spin Chern number for a specific spin state is nonzero. The second one is the mirror-reflection symmetry breaking, based on which the Dirac cones generated at the BZ corners will open a bandgap. This bandgap, projected by valley-polarized states, is also topologically nontrivial. Recently, the coupling combination of pseudo-spin and valley states in photonic systems further enables the unidirectional edge propagation, the one-way Klein tunneling, and the topological corner polarization of electromagnetic waves[33-37].

The higher-order topological insulators (HOTIs) is a new class of TIs which have been theoretically proposed in quantized electronic multipole systems[38-47]. The topological phases of HOTIs are characterized by the bulk polarizations but not the integer topological invariants, going beyond the conversional bulk-boundary correspondence. The 2D HOTIs do not exhibit the 1D gapless edge states, but instead the zero-dimensional (0D) topological states on the "boundaries of boundaries". So far, the higher-order topological phases, stemming from the quantization of quadrupole moments, were experimentally realized in mechanical[48] and microwave systems[49] and electrical circuits[50] with ingeniously dimerized hopping, described by positive and negative couplings. Very recently, based on the skillful modulation of inter and intra-cell couplings, the higher-order topological states, expressing as the gapped 1D edge states and the topologically protected nontrivial in-gap 0D corner states, have been realized in conversional optical, acoustic and elastic systems[51-58]. Here, a pseudospin-valley coupling mechanism will be developed to yield the multiple edge and corner states of elastic waves which are topologically-protected in a well-controlled manner.

**Results**

The phononic analogues of topological pseudospin and valley phases are uniformly realized on a simple elastic phononic crystal (PnC). The elastic PnC is fabricated by cutting hexagonal blocks from an acrylic plate. Cylindrical nickel-plated neodymium magnets are attached at nodes of the elastic PnC, being as additional masses (Fig. 1). For the perfect elastic PnC without any perturbation, the gapless Hamiltonian will form the Dirac cones at the BZ corners (K and K') which are essentially protected by the mirror-reflection symmetry of the unit cell [27]. The mirror-reflection symmetry can be broken by varying the additional masses. The additional masses on the A and B nodes are defined as $m_A=m_0+\Delta m$ and $m_B=m_0-\Delta m$. The $m_0$ is the mass of magnet in the perfect elastic PnC without any perturbation. When $\Delta m \neq 0$, the Dirac cones open to a bandgap, giving rise to a valley-polarized edge state[25,27,31]. The lattice deformation emulating the spin degree of freedom can be induced by the perturbation of the composite unit cell consisting of three primitive unit cells. The intra-cell beam length is defined as $l_{\text{intra}}=(1-\Delta\gamma)l$ and at the same time, the inter-cell beam length is defined as $l_{\text{inter}}=(1+2\Delta\gamma)l$. The length of the acrylic beam of the perfect elastic PnC without any perturbation is $l$=15mm. When the composite unit cell is expanded ($\Delta\gamma<0$) or shrunk ($\Delta\gamma>0$), the double Dirac cone at the BZ center (Γ) which is folded from the Dirac cones at the BZ corners (K and K'), will be lifted. The inversion of the $p$ and $d$ modes (similar to the $p$ and $d$ orbitals of electrons) at the BZ center can be induced by the lattice deformation with the opposite $\Delta\gamma$ (see Fig. S1 in the Supplemental Material for $\Delta\gamma$=0.164 and $\Delta\gamma=-0.2$). The band gap of a composite unit cell with both nonzero $\Delta m$ and $\Delta\gamma$ is wider than that of a unit cell with only one nonzero $\Delta m$ or $\Delta\gamma$ (see Fig. S2 in the Supplemental Material). Composite unit cells with ($\Delta\gamma$=0.164, $\Delta m$=0.5) and ($\Delta\gamma=-0.2$, $\Delta m=-0.5$) are depicted in Figs. 1a and 1b.

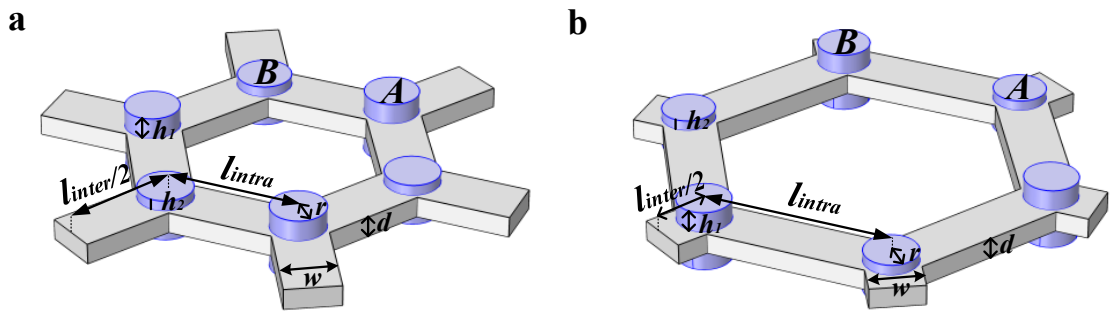

Fig. 1 | **a**, Composite unit cells with $\Delta\gamma$=0.164, $\Delta m$=0.5. **b**, The composite unit cells with $\Delta\gamma=-0.2$, $\Delta m=-0.5$. The parameters are $w$=5.02mm, $d$=1.98mm, $r$=2.51mm, $h_1$=2.0mm and $h_2$=1.0mm.

As this elastic PnC can be considered as a thin plate in a long wavelength, the in-plane polarization is loosely coupled with the out-of-plane polarization characterized by the parabolic dispersion. Therefore, the out-of-plane bands, uncoupled from the in-plane bands (gray dots in the bandgap in Fig.2), can be ignored. The dispersion of a supercell consisting of twelve shrunk unit cells ($\Delta\gamma = 0.164$, $\Delta m=0$) in the upper domain and thirteen expanded unit cells ($\Delta\gamma = -0.2$, $\Delta m=0$) in the lower domain, separated by an armchair interface, is conducted to evaluate the edge states. For this supercell with the lattice deformation but keeping the mirror-reflection symmetry, two nearly gapless edge states emerge inside the bandgap, as shown in Fig. 2a. There is a very small bandgap between the two elastic edge states at $k_x$=0 due to the breaking of the $C_{6v}$ symmetry at the armchair interface[59]. The $p$ and $d$ modes (Fig. S1) hybridize to a symmetric component $S = (p_x + d_{x^2-y^2})/\sqrt{2}$ and an anti-symmetric component $A = (p_y + d_{xy})/\sqrt{2}$. Elastic pseudospins of the edge modes are represented as S+iA/S−iA by hybridizing the symmetric component (S) and the anti-symmetric component (A) illustrated in the right panel of Fig. 2a. The band dispersion of a supercell consisting of twelve primitive unit cells with $\Delta\gamma$=0 and $\Delta m$=−0.5 at the upper domain and other twelve primitive unit cells with $\Delta\gamma$=0 and $\Delta m$=0.5 at the lower domain, separated by a zigzag interface, is depicted in Fig. 2b. For this supercell with the mirror-reflection symmetry breaking but without any lattice deformation, a gapless edge state emerges inside the bandgap, as shown in Fig. 2b. This gapless edge state, localized at the interface between two different unit cells, supports a valley polarized edge propagation[27].

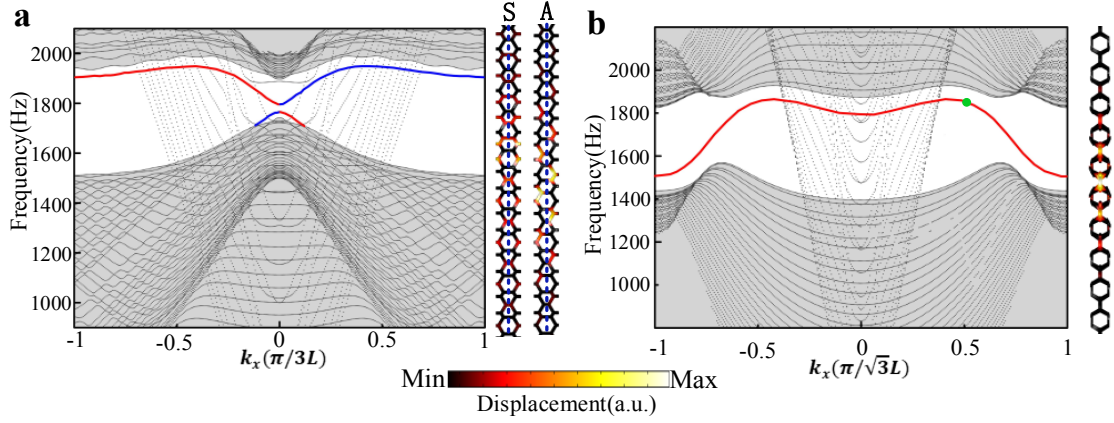

Fig. 2 | **a**, Band dispersion of a ribbon configuration with lattice deformation only. The simulated displacement field profiles of partial supercell in the (out-of-plane) z direction at $k_x$=0 is displayed at the right panel. **b**, Band dispersion of a ribbon configuration with symmetry breaking only. The simulated displacement field profile of the green node is displayed at the right panel. The gray shaded parts are bulk bands. The gray dots in the bandgap are the in-plane bands.

In this paper, the lattice deformation and the mirror-reflection symmetry breaking will be simultaneously introduced in our elastic PnC. The photograph of the fabricated elastic PnC sample is illustrated in Fig. 3a. The horizontal blue line denotes the armchair interface separating the phononic crystals with $\Delta\gamma$=0.164 (top) and $\Delta\gamma$=−0.2 (bottom). The vertical red line denotes the zigzag interface separating the phononic crystals with $\Delta m$=−0.5 (left) and $\Delta m$=0.5 (right). There are four types pseudospin–valley coupled unit cells with ($\Delta m$=0.5, $\Delta\gamma$=0.164), ($\Delta m$=−0.5, $\Delta\gamma$=0.164), ($\Delta m$=−0.5, $\Delta\gamma$=−0.2) and ($\Delta m$=0.5, $\Delta\gamma$=−0.2). They respectively locate at the upper-right, the upper-left, the lower-left and the lower-right sections of the fabricated sample, as shown in Fig. 3a. Among these sections, there are four different domain walls (DWs) labelled as DW1, DW2, DW3, and DW4. The absolute values of $\Delta m$ in supercells with DW4 and DW1 are the same, namely, pseudospin-valley coupling in the supercells with DW4 and DW1 are the same, thus the edge state of DW4 is the same as that of DW1. The absolute values of $\Delta\gamma$ in supercells with DW2 and DW3 are different, namely, the pseudospin-valley coupling in the supercells with DW2 and DW3 are different, thus the frequencies of edge states of DW2 and DW3 are different. Band dispersions of supercells with DW1, DW2 and DW3 are indicated in Figs. 3b-3d, respectively. As displayed in Figs. 3b-3d, these edge states are gapped. It is a unique characteristic of the pseudospin-valley coupled phononic TI. For a supercell in Fig. 2a (Fig. 2b), $\Delta\gamma$ ($\Delta m$) changes its sign across the interface with $\Delta m=0$ ($\Delta\gamma=0$), leading to gapless edge states. For a supercell in Fig. 3b, $\Delta\gamma$ changes its sign across the DW1, leading to topological phases transition. However, $\Delta m$ is nonzero and keeps the same topological phases across the DW1, resulting in a complete bandgap without any edge state. Thus, due to the coupling of $\Delta\gamma$ and $\Delta m$, the edge state in Fig.3b is gapped. Likewise, the edge states of DW2 and DW3 in Figs.3c-3d are gapped. The evolution of edge states from gapless to gapped ones with the increasing of pseudospin-valley coupling is shown Fig. S3 of the Supplemental Material.

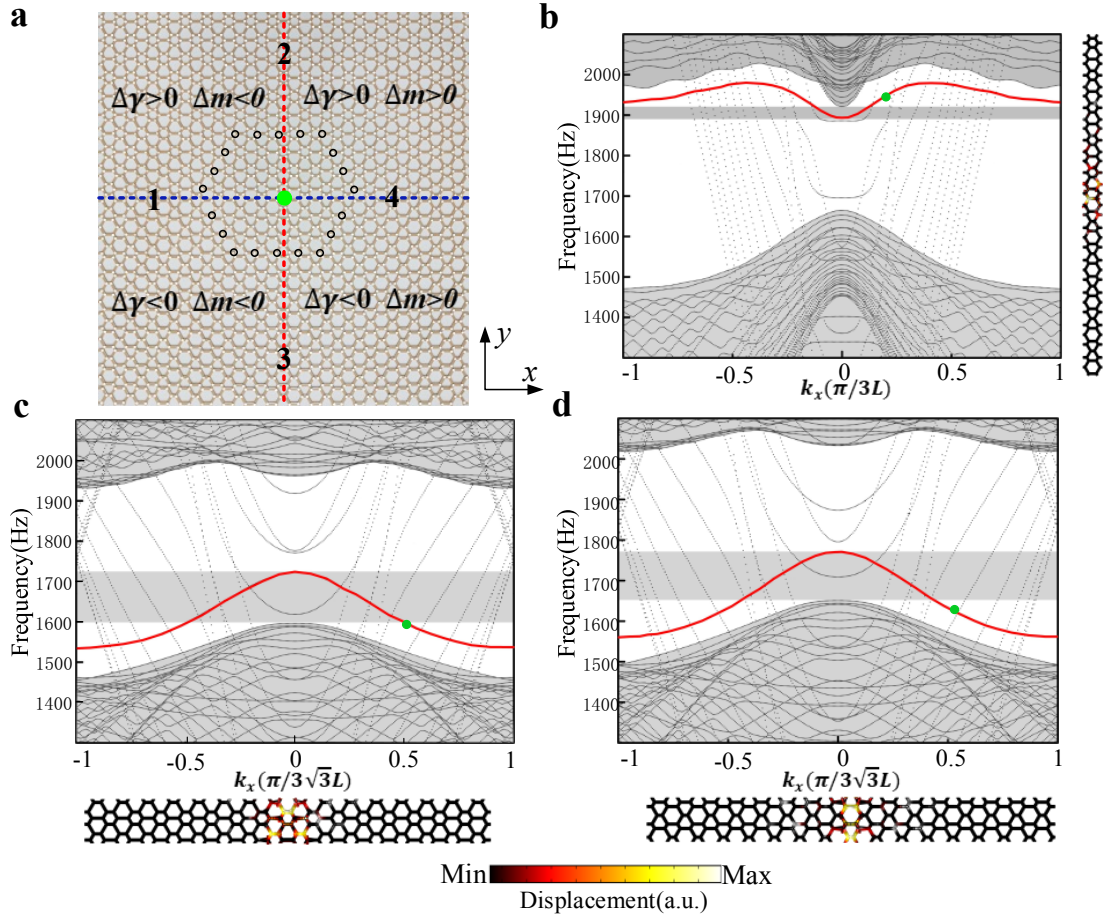

Fig. 3 | **a**, The photograph of the fabricated elastic PnC sample. The black circles indicate locations of defects. **b-d**, Band dispersions of supercells with DW1, DW2 and DW3. The simulated displacement field profiles of green nodes are displayed. Gray shaded parts are bulk bands. Gray dots in the bandgap are in-plane bands.

Four elastic PnC samples without any defects are firstly considered. The first one is a sample with DW1, the second one is a sample with DW2, the third one is a sample with DW3, and the last one is a sample with four types of DWs. Numerically evaluated eigenfrequencies of four rectangular samples are presented in Fig. 4a. There are only bulk modes (black, green and orange circle dots) and gapped edge modes (black, green and orange square points) in the first three samples, as expected. The frequency of the edge mode of DW1 is from 1890.6Hz to 1907.7Hz which matches with the edge state shown in Fig. 3b. The frequency of the edge mode of DW2 is from 1634.8Hz to 1689.5Hz and it is from 1663.8 Hz to 1755Hz for DW3, which match with the edge states of DW2 and DW3 shown in Figs. 3c and 3d. A corner mode (marked by the red star) inside the gap of edge states (marked by red square points) is observed in the fourth sample. To confirm that the corner state here is topologically protected, we deliberately introduce defects by removing magnets at both sides of 20 nodes (marked by black circles in Fig. 3a). Numerical results of eigenfrequencies marked by blue points in Fig. 3a show that the edge eigenmodes (the blue square points) and the corner mode (surround by the blue star) are well confined at the same frequencies, when compared with the fourth sample without defects. It is a convincing evidence that the edge states and the corner state of our pseudospin-valley coupled phononic TI exhibit a strong robustness to the moderate defects.

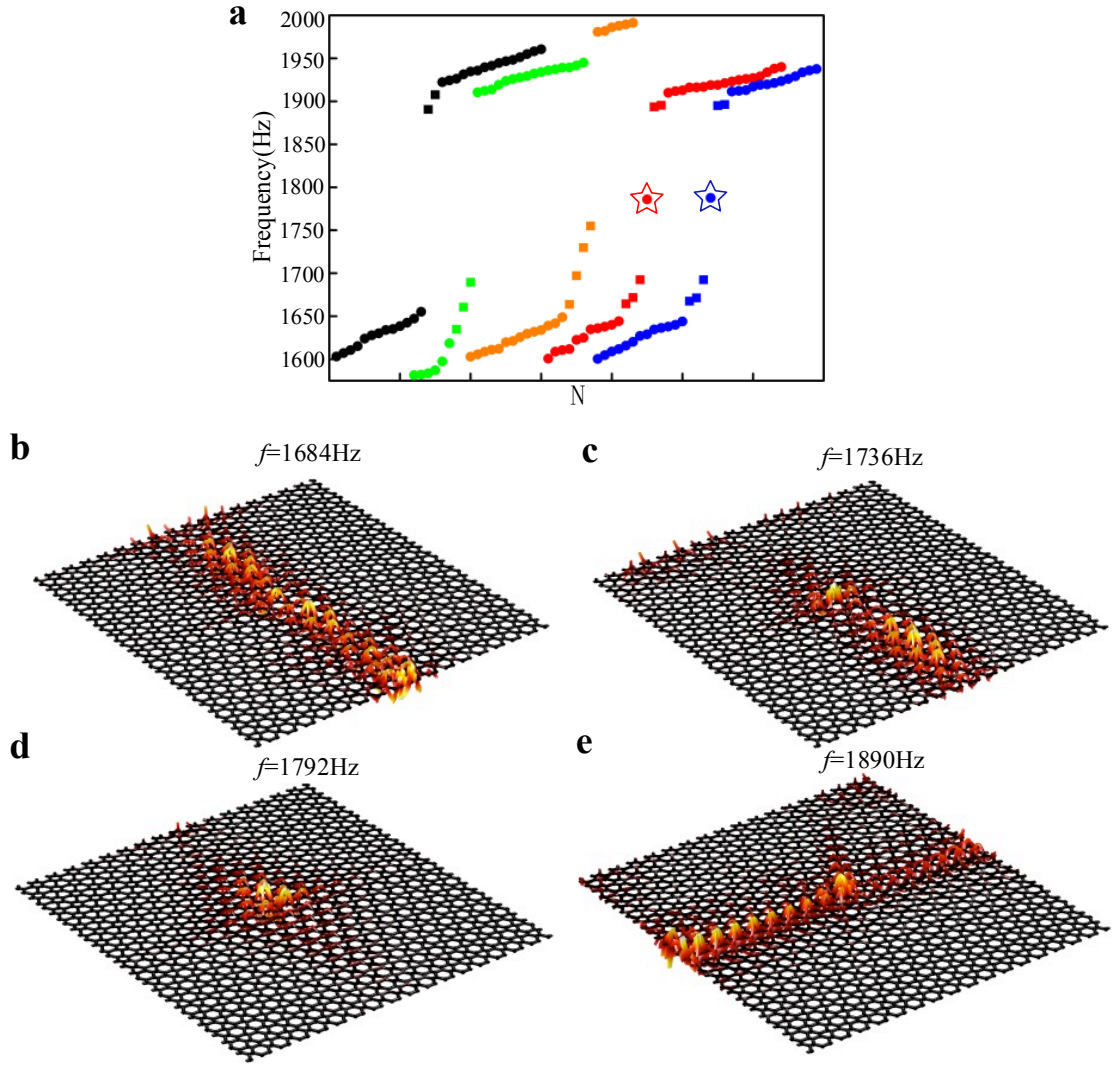

Fig. 4 | **a**, Numerically calculated eigenfrequencies of the samples with DW1 (black points), DW2 (green points), DW3 (orange points), four types of DWs without defect (red points) and four types of DWs with defect (blue points), respectively. The square points denote gapped edge modes in five different rectangular samples. The corner modes are highlighted by stars. **b-e**, The simulated displacement field profiles at frequencies 1684, 1736, 1792 and 1890Hz, respectively.

By exciting the fourth elastic PnC sample with a vertical force at the crossing point (denoted by the green dot in Fig. 3a), simulated displacement field profiles at different frequencies can be obtained (Figs. 4b-4e). In Fig. 4b, when the frequency is 1684Hz, the elastic wave energy is well localized at the DW2 and DW3 of the elastic PnC sample with four types of DWs. Moreover, when the frequency is 1736Hz which is within the edge band of DW3, but outside of the edge band of DW2, the elastic wave energy is only localized at the DW3 (Fig. 4c). When the frequency increases to 1792Hz between the edge states of DW3 and DW1, the topological corner state is excited (Fig. 4d). The spatial exponential attenuation of the displacement field intensity clearly characterizes the topological corner mode. The gapped edge states in DW1 and DW4 are manifested at the frequency of 1890Hz (Fig. 4e).

The measured edge transmission spectra of the fabricated elastic PnC sample (Fig. 3a) are presented in Fig. 5a. Within the gap of the edge band of DW1, the spectrum (the blue curve) exhibits a very low transmission. On the contrary, a high peck, induced by edge state emerging in the frequency domain from 1892.3Hz to 1921Hz, is observed at 1990Hz. This result provides a strong evidence of the edge state of DW1. The edge transmission spectra of edge states of DW2 and DW3 exhibit simultaneously high transmissions from 1650Hz to 1721Hz. However, from 1722Hz to 1770Hz, the edge transmission spectrum of DW2 (the black curve) decreases sharply, while the edge transmission spectrum of DW3 (the red curve) is still high. This phenomenon is consistent with the edge bands shown in the simulations of Fig. 3c-3d. The corner transmission spectra of the fabricated elastic PnC sample (Fig. 3a) are presented in Fig. 5b. For the corner transmission spectrum without defects (the black curve), the peak is observed in the gap region among the edge bands. The measured corner transmission spectrum with defects by removing magnets at both sides of 20 nodes (marked by black circles in Fig. 3a) is presented by the red curve in Fig. 5b. It indicates that the high peak is around 1800Hz, which is consistent with the black curve. This phenomenon demonstrates that the corner state is topologically protected and exhibit the strong robustness against the moderate defects. All measurements were carried out by a scanning laser Doppler vibrometer (LV-S01), seeing Methods in the Supplemental Material.

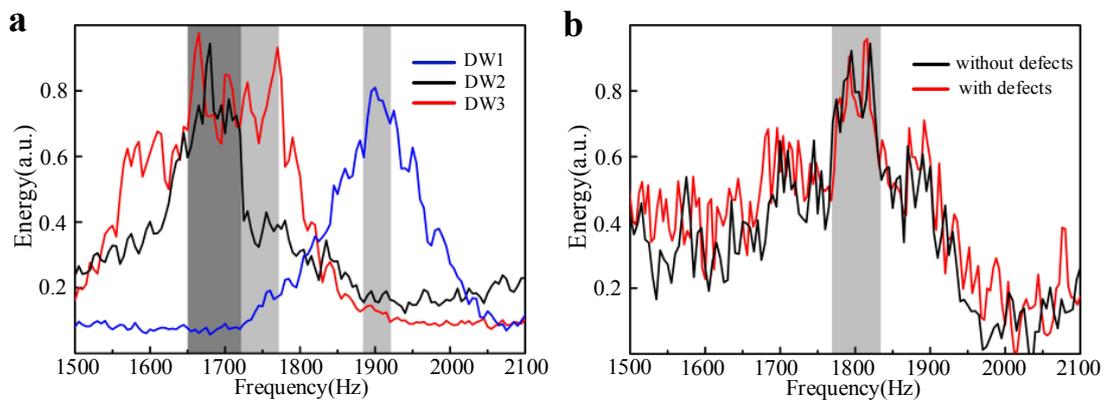

Fig. 5 | **a**, Measured edge transmission spectra of the fabricated elastic PnC sample. **b**, Measured corner transmission spectra of the fabricated elastic PnC samples with defects (red curve) and without defects (black curve).

In general, this research proposes a pseudospin-valley-coupled phononic TI in a continuous elastic system which simultaneously possesses the gapped edge states and the topological corner state. By manipulating the distance between neighboring magnets, the pseudospin-orbit coupling edge states are induced by the lattice deformation. Besides, the valley-polarized gapless edge state can be obtained by the mirror-reflection symmetry breaking. When the lattice deformation and the mirror-reflection symmetry breaking are introduced simultaneously, the topological gapless edge states evolve to be gapped and the topological corner state emerges. As the frequencies of topological edge states in the distinct DWs are different, the elastic wave energy can be well localized at different DWs in a well-control means. For the corner state, the elastic wave energy is well concentrated at the intersection of the four types of DWs. Our work provides a reconfigurable platform for the well-control of the multiple edge states and the corner state as the pseudospin-valley coupling is tunable. The potential applications of our pseudospin-valley coupled phononic topological insulator in the elastic energy recovery and the highly sensitive sensing is also promising.

## Acknowledgments

The paper is supported by the Foundation for Innovative Research Groups of the National Natural Science Foundation of China (Grant No. 51621004) and the Joint Fund of Ministry of Education for Equipment Pre-Research (6141A02033216).

## Author contributions

H.Y. and B.Z. designed the experiments. H.Y. and L. performed the experiments. H.Y. and S.J. carried out the numerical simulations. H.Y., and B.Z. wrote the manuscript. B.Z supervised the project. All the authors contributed to the analyses of results.

## Competing interests

The authors declare no competing interests.

# References

1  Poo, Y., Wu, R.-x., Lin, Z., Yang, Y. & Chan, C. Experimental Realization of Self-Guiding Unidirectional Electromagnetic Edge States. *Physical Review Letters* **106**, 093903 (2011).

2  Skirlo, S. A., Lu, L. & Soljačić, M. Multimode One-Way Waveguides of Large Chern Numbers. *Physical Review Letters* **113**, 113904 (2014).

3  Bahari, B. *et al.* Nonreciprocal lasing in topological cavities of arbitrary geometries. *Science* **358**, 636-640 (2017).

4  Rechtsman, M. C. *et al.* Photonic Floquet topological insulators. *Nature* **496**, 196-200 (2013).

5  Maczewsky, L. J., Zeuner, J. M., Nolte, S. & Szameit, A. Observation of photonic anomalous Floquet topological insulators. *Nature Communications* **8**, 13756 (2017).

6  Mukherjee, S. *et al.* Experimental observation of anomalous topological edge modes in a slowly driven photonic lattice. *Nature Communications* **8**, 13918 (2017).

7  Leykam, D. & Chong, Y. D. Edge Solitons in Nonlinear-Photonic Topological Insulators. *Physical Review Letters* **117**, 143901 (2016).

8  He, C. *et al.* Topological phononic states of underwater sound based on coupled ring resonators. *Applied Physics Letters* **108**, 031904 (2016).

9  Cheng, X. J. *et al.* Robust reconfigurable electromagnetic pathways within a photonic topological insulator. *Nature Materials* **15**, 542-548 (2016).

10 Khanikaev, A. B. *et al.* Photonic topological insulators. *Nature Materials* **12**, 233-239 (2013).

11 Ma, T., Khanikaev, A. B., Mousavi, S. H. & Shvets, G. Guiding Electromagnetic Waves around Sharp Corners: Topologically Protected Photonic Transport in Metawaveguides. *Physical Review Letters* **114**, 127401 (2015).

12 Zhu, X. *et al.* Topological transitions in continuously deformed photonic crystals. *Physical Review B* **97**, 085148 (2018).

13 Wu, L.-H. & Hu, X. Scheme for Achieving a Topological Photonic Crystal by Using Dielectric Material. *Physical Review Letters* **114**, 223901 (2015).

14 Yang, Y. *et al.* Visualization of a Unidirectional Electromagnetic Waveguide Using Topological Photonic Crystals Made of Dielectric Materials. *Physical Review*

*Letters* **120**, 217401 (2018).

15    Dong, J.-W., Chen, X.-D., Zhu, H., Wang, Y. & Zhang, X. Valley photonic crystals for control of spin and topology. *Nature materials* **16**, 298-302 (2017).

16    Gao, F. *et al.* Topologically protected refraction of robust kink states in valley photonic crystals. *Nature Physics* **14**, 140–144 (2018).

17    Chen, X. D. *et al.* Tunable Electromagnetic Flow Control in Valley Photonic Crystal Waveguides. *Physical Review Applied* **10**, 044002 (2018).

18    Chen, X.-D., Zhao, F.-L., Chen, M. & Dong, J.-W. Valley-contrasting physics in all-dielectric photonic crystals: Orbital angular momentum and topological propagation. *Physical Review B* **96**, 020202 (2017).

19    Gao, Z. *et al.* Valley surface-wave photonic crystal and its bulk/edge transport. *Physical* Review *B* **96**, 201402 (2017).

20    Noh, J., Huang, S., Chen, K. P. & Rechtsman, M. C. Observation of Photonic Topological Valley Hall Edge States. *Physical Review Letters* **120**, 063902 (2018).

21    Mousavi, S. H., Khanikaev, A. B. & Wang, Z. Topologically protected elastic waves in phononic metamaterials. *Nature Communications* **6**, 8682 (2015).

22    He, C. *et al.* Acoustic topological insulator and robust one-way sound transport. *Nature Physics* **12**, 1124-1129 (2016).

23    Foehr, A., Bilal, O. R., Huber, S. D. & Daraio, C. Spiral-based phononic plates: From wave beaming to topological insulators. *Physical review letters* **120**, 205501 (2018).

24    Zhang, Z. *et al.* Topological Creation of Acoustic Pseudospin Multipoles in a Flow-Free Symmetry-Broken Metamaterial Lattice. *Physical Review Letters* **118**, 084303 (2017).

25    Yan, M. *et al.* On-chip valley topological materials for elastic wave manipulation. *Nature Materials* **17**, 993-998 (2018).

26    Lu, J. *et al.* Observation of topological valley transport of sound in sonic crystals. *Nature Physics* **13**, 369-374 (2017).

27    Vila, J., Pal, R. K. & Ruzzene, M. Observation of topological valley modes in an elastic hexagonal lattice. *Physical Review B* **96**, 134307 (2017).

28    Brendel, C., Peano, V., Painter, O. J. & Marquardt, F. Pseudomagnetic fields for sound at the nanoscale. *Proceedings of the National Academy of Sciences* **114**, E3390-E3395 (2017).

29  Li, S., Zhao, D., Niu, H., Zhu, X. & Zang, J. Observation of elastic topological states in soft materials. *Nature communications* **9**, 1370 (2018).

30  Yu, S. Y. *et al.* Elastic pseudospin transport for integratable topological phononic circuits. *Nature Communications* **9**, 3072 (2018).

31  Liu, T.-W. & Semperlotti, F. Tunable Acoustic Valley–Hall Edge States in Reconfigurable Phononic Elastic Waveguides. *Physical Review Applied* **9**, 014001 (2018).

32  Miniaci, M., Pal, R. K., Morvan, B. & Ruzzene, M. Experimental Observation of Topologically Protected Helical Edge Modes in Patterned Elastic Plates. *Physical Review X* **8**, 031074 (2018).

33  Kang, Y. H., Ni, X., Cheng, X. J., Khanikaev, A. B. & Genack, A. Z. Pseudo-spin-valley coupled edge states in a photonic topological insulator. *Nature Communications* **9**, 3029 (2018).

34  Ma, T. & Shvets, G. Scattering-free edge states between heterogeneous photonic topological insulators. *Physical Review B* **95**, 165102 (2017).

35  Ni, X. *et al.* Spin-and valley-polarized one-way Klein tunneling in photonic topological insulators. *Science Advances* **4**, eaap8802 (2018).

36  Xue, H. *et al.* Spin-valley-controlled photonic topological insulator. *arXiv preprint arXiv:1811.00393* (2018).

37  Yang, Y. *et al.* Gapped Topological Kink States and Topological Corner States in Graphene. *arXiv preprint arXiv:1903.01816* (2019).

38  Benalcazar, W. A., Bernevig, B. A. & Hughes, T. L. Quantized electric multipole insulators. *Science* **357**, 61-66 (2017).

39  Song, Z. D., Fang, Z. & Fang, C. (d-2)-Dimensional Edge States of Rotation Symmetry Protected Topological States. *Physical Review Letters* **119**, 246402 (2017).

40  Langbehn, J., Peng, Y., Trifunovic, L., von Oppen, F. & Brouwer, P. W. Reflection-Symmetric Second-Order Topological Insulators and Superconductors. *Physical Review Letters* **119**, 246401 (2017).

41  Ezawa, M. Higher-Order Topological Insulators and Semimetals on the Breathing Kagome and Pyrochlore Lattices. *Physical Review Letters* **120**, 026801 (2018).

42  Lin, M. & Hughes, T. L. Topological quadrupolar semimetals. *Physical Review B* **98**, 241103 (2018).

43  Li, L. H., Umer, M. & Gong, J. B. Direct prediction of corner state configurations from edge winding numbers in two- and three-dimensional chiral-symmetric lattice systems. *Physical Review B* **98**, 205422 (2018).

44  Khalaf, E. Higher-order topological insulators and superconductors protected by inversion symmetry. *Physical Review B* **97**, 205136 (2018).

45  Ezawa, M. Higher-order topological electric circuits and topological corner resonance on the breathing kagome and pyrochlore lattices. *Physical Review B* **98**, 201402 (2018).

46  Ezawa, M. Minimal models for Wannier-type higher-order topological insulators and phosphorene. *Physical Review B* **98**, 045125 (2018).

47  Benalcazar, W. A., Bernevig, B. A. & Hughes, T. L. Electric multipole moments, topological multipole moment pumping, and chiral hinge states in crystalline insulators. *Physical Review B* **96**, 245115 (2017).

48  Serra-Garcia, M. *et al.* Observation of a phononic quadrupole topological insulator. *Nature* **555**, 342–345 (2018).

49  Peterson, C. W., Benalcazar, W. A., Hughes, T. L. & Bahl, G. A quantized microwave quadrupole insulator with topologically protected corner states. *Nature* **555**, 346–350 (2018).

50  Imhof, S. *et al.* Topolectrical-circuit realization of topological corner modes. *Nature Physics* **14**, 925-929, (2018).

51  Noh, J. *et al.* Topological protection of photonic mid-gap defect modes. *Nature Photonics* **12**, 408-415, (2018).

52  Zhang, X. *et al.* Second-order topology and multidimensional topological transitions in sonic crystals. *Nature Physics* 15 (6), 582-88 (2019)..

53  Zhang, X. *et al.* Acoustic Hierarchical Topological Insulators. *arXiv preprint arXiv:1811.05514* (2018).

54  Xie, B. Y. *et al.* Second-order photonic topological insulator with corner states. *Physical Review B* **98**, doi:ARTN 205147

10.1103/PhysRevB.98.205147 (2018).

55  Ni, X., Weiner, M., Alu, A. & Khanikaev, A. B. Observation of higher-order topological acoustic states protected by generalized chiral symmetry. *Nature Materials* **18**, 113-20 (2019).

56  Xue, H. R., Yang, Y. H., Gao, F., Chong, Y. D. & Zhang, B. L. Acoustic higher-

order topological insulator on a kagome lattice. *Nature Materials* **18**, 108-112 (2019).

57  Zhang, Z. W., Lopez, M. R., Cheng, Y., Liu, X. J. & Christensen, J. Non-Hermitian Sonic Second-Order Topological Insulator. *Physical Review Letters* **122,** 195501 (2019).

58  Fan, H. Y., Xia, B. Z., Tong, L., Meng, S. J. & Yu, D. J. Elastic Higher-Order Topological Insulator with Topologically Protected Corner States. *Physical Review Letters* **122**, 204301 (2019).

59  Deng, Y., Ge, H., Tian, Y., Lu, M. & Jing, Y. Observation of zone folding induced acoustic topological insulators and the role of spin-mixing defects. *Physical Review B* **96**, 184305 (2017).